\documentclass[aps,prx,reprint,superscriptaddress,amsmath,amssymb]{revtex4-2}

\usepackage{adjustbox}
\usepackage{tikz}
\usetikzlibrary {graphs,graphs.standard, positioning,backgrounds}
\usepackage{graphicx}% Include figure files
\usepackage{dcolumn}% Align table columns on decimal point
\usepackage{bm}% bold math
\usepackage{amsthm,amsfonts}
\usepackage{color}
\usepackage{hyperref}% add hypertext capabilities

\usepackage{multirow} % multi row in table
\usepackage[ruled]{algorithm2e}

\newcommand{\comment}[1]{}

\DeclareMathOperator{\contract}{contract}
\DeclareMathOperator{\sing}{sing}
\DeclareMathOperator{\tr}{Tr}
\DeclareMathOperator{\diag}{diag}

\newtheorem{theorem}{Theorem}[section]  %set theorem counter
\newtheorem{lemma}[theorem]{Lemma}  %set lemma counter
\begin{document}

\title{Simulation of the Ring-exchange Models with Projected Entangled Pair States}

\author{Chao Wang}
\affiliation{CAS Key Laboratory of Quantum Information, University of Science and Technology of China, Hefei 230026, Anhui, China}
\affiliation{Synergetic Innovation Center of Quantum Information and Quantum Physics, University of Science and Technology of China, Hefei, 230026, China}
\author{Shaojun Dong}
\affiliation{CAS Key Laboratory of Quantum Information, University of Science and Technology of China, Hefei 230026, Anhui, China}
\affiliation{Synergetic Innovation Center of Quantum Information and Quantum Physics, University of Science and Technology of China, Hefei, 230026, China}
\author{Yongjian Han}%
\email{smhan@ustc.edu.cn}
\affiliation{CAS Key Laboratory of Quantum Information, University of Science and Technology of China, Hefei 230026, Anhui, China}
\affiliation{Synergetic Innovation Center of Quantum Information and Quantum Physics, University of Science and Technology of China, Hefei, 230026, China}
\author{Lixin He}
\email{helx@ustc.edu.cn}
\affiliation{CAS Key Laboratory of Quantum Information, University of Science and Technology of China, Hefei 230026, Anhui, China}
\affiliation{Synergetic Innovation Center of Quantum Information and Quantum Physics, University of Science and Technology of China, Hefei, 230026, China}

\date{\today}% It is always \today, today,
             %  but any date may be explicitly specified

\begin{abstract}
Algorithms to simulate the ring-exchange models using the projected entangled pair states (PEPS) are developed.
We generalize the imaginary time evolution (ITE) method to optimize PEPS wave functions for the models with ring-exchange interactions.
We compare the effects of different approximations to the environment.
To understand the numerical instability during the optimization, we introduce the ``singularity'' of a PEPS and develop a regulation procedure that can effectively reduce the singularity of a PEPS.
We benchmark our method with the toric code model, and obtain extremely accurate ground state energies and
topological entanglement entropy. We also benchmark our method with the two-dimensional cyclic ring exchange model, and find that the ground state has a strong vector chiral order. The algorithms can be a powerful tool to investigate the models with ring interactions. The methods developed in this work, e.g., the regularization process to reduce the singularity can also be applied to other models.

\end{abstract}

\maketitle

\section{\label{sec:into}Introduction}

Recently, the models with ring-exchange interactions have attracted growing attention.
The ring-exchange interactions were first introduced as the higher order perturbation terms of
the Hubbard model near half-filling\cite{MacDonald88,Chubukov92,Kolezhuk98,Brehmer99,
Muller02,Lauchli03,Gritsev04,Delannoy05,Song06,Delannoy09}.
It has been shown that the ring-exchange interactions may play important roles in many materials \cite{Katanin02,Calzado03,Toader05,Delannoy09}.
The ring-exchange terms %in Bosonic model
can stabilize the uncondensed Boson liquid \cite{Motrunich07,Sheng08,Block11,Mishmash11,Huerga14},
which helps to reveal the physics of strange metal in the High-Tc superconductors\cite{Galitski05}.
The J-Q model, which includes the ring-exchange interactions, has been used
\cite{Sandvik07,Sandvik10} to demonstrate the deconfined quantum critical point\cite{Senthil04}.
The ring-exchange interactions are also crucial in the lattice-gauge models\cite{Senthil00,Moessner01,Kitaev03} to exhibit topological phase transition. Recently,
the ring-exchange interactions have been realized experimentally in the cold atom systems\cite{Dai17,Bohrdt20}.

The ring-exchange models have been studied by various numerical methods, including the
exact diagonalization (ED)\cite{Roux05,Song06,Sheng08,Larsen19}, the
quantum Monte Carlo (QMC)\cite{Evertz10}, and the density matrix renormalization group (DMRG)~\cite{White92} methods. However, those methods all have some limitations. For example, the ED method can only treat rather small systems,
whereas the QMC may suffer from the sign problems in the fermionic and frustrated systems\cite{Henelius00,Sandvik10}.
The DMRG method can only treat the quasi-one-dimensional systems \cite{Roux05,Sheng08,Ramos14}.

The recent developed tensor network states (TNS)
methods\cite{Orus14,Vidal03,Verstraete04,Vidal08,Xie14}, including the projected entangled states (PEPS) method, are very promising to study the two-dimensional many-particle systems,
which has been successfully applied to various models, such as the $J_1$-$J_2$ model\cite{Liu18}, kagome Heisenberg model\cite{Liao17}, and the $t$-$J$ model\cite{Coboz14,Dong20}, etc.
However, so far, the study of ring-exchange model via PEPS is still very rare\cite{Crone20}.
The ring-exchange models involve 4-site interactions,
which are significantly more complicate than the bond interaction models (e.g. the Heisenberg model) for the PEPS.

In principle, the ground state PEPS wave functions of the ring-exchange models
can be optimized via the gradient optimization methods~\cite{Crone20,Liu17}.
However, the gradient optimization starting from a random PEPS may suffer from local minima.
The imaginary time evolution method (ITE) may be more efficient and may effectively avoid the local minima. The results of ITE can be used as the starting wave functions for further gradient optimization~\cite{Liu17}. However, so far no effective ITE algorithm has been developed for the ring-exchange models, because of the complication of the 4-site interactions.
In this work, we develop an efficient ITE update algorithm to optimize the PEPS for the
ring-exchange models. Combined with the stochastic gradient optimization methods~\cite{Liu17},
it provides a powerful tool to study the ring-exchange models.

The paper is organized as follows. In Sec.\ref{sec:meth}, we introduce the algorithms to optimize the PEPS for the ring-exchange models. After introduce a general algorithm of ITE for the ring-exchange model, we discuss the approximations of the environment in Sec.\ref{sec:env}. In Sec.\ref{sec:sing} we discuss how to control the ``singularity'' of the PEPS, to enhance the numerical stability during the ITE.
In Sec.\ref{sec:mod1}, we benchmark the algorithms on the toric code model, in which we calculate the ground state energies and the topological entanglement entropy of the model.
In Sec.\ref{sec:mod2}, we apply the algorithms to study the cyclic ring-exchange model.
We summarize in Sec.\ref{sec:con}.

\section{\label{sec:meth}Method}

We consider a general Hamiltonian with ring-exchange interactions on the square lattices,
\begin{equation}
    H=\sum_i H_{1,i}+\sum_{i,j}H_{2,i,j}+\sum_p H_{\square,p} \, ,
\end{equation}
where $H_{1,i}$, $H_{2,i,j}$ denote the one-site and two-site interaction terms respectively.
$H_{\square,p}$ which act on plaquettes $p$, denote the ring-exchange interaction terms.

We solve the ground state of the above Hamiltonian using the PEPS method\cite{Verstraete04}.
A PEPS on an $L_1\times L_2$ square lattice can be written as,
\begin{equation} |\Psi\rangle=\prod_{i=1,j=1}^{L_1,L_2}\Big[T^{[i,j]}_{l_{ij}r_{ij}u_{ij}d_{ij}s_{ij}}
\delta_{r_{ij}l_{i,j+1}}\delta_{d_{ij}u_{i+1,j}}\Big]\bigotimes_{i=1,j=1}^{L_1,L_2}|s_{ij}\rangle
\label{eq:TNS}
\end{equation}
where Einstein's sum rule is assumed. In this equation,
$T^{[i,j]}_{l_{ij}r_{ij}u_{ij}d_{ij}s_{ij}}$ is a rank-5 complex tensor at site $(i,j)$ which has four auxiliary indices $l_{ij},r_{ij},u_{ij},d_{ij}$ and a physical index $s_{ij}$.
Each auxiliary indices are summed from 1 to the bond dimension $D$,
and each physical index is summed from 1 to the physical dimension $d$.

The ground state $|\Psi_{\rm GS}\rangle$ of Hamiltonian $H$ in the form of PEPS
can be optimized by ITE algorithm~\cite{Vidal04},
\begin{align}
	|\Psi_{\rm GS}\rangle&= \lim_{\tau\rightarrow \infty}e^{-H\tau}|\Psi_0\rangle \nonumber\\
	&=\lim_{N\rightarrow \infty}(\prod_{i=1}^ne^{-\Delta\tau H_i}\prod_{i=n}^1e^{-\Delta\tau H_i})^N|\Psi_0\rangle
\end{align}
where $|\Psi_0\rangle$ is a randomly chosen initial PEPS.
The ITE is divided into small time steps, i.e.,
$\Delta\tau \ll 1$, which can be well approximated by
the products of a series of local operators $H_i$ using the
Trotter expansion~\cite{Hatano05}.

The key ingredient of the ITE algorithm
is to efficiently approximate the $(n+1)$-th step wave function $|\Psi_{n+1}\rangle$=
$e^{-\Delta \tau H_{i}}|\Psi_n\rangle$ by a PEPS, which we call a local update process.
We regard the sites, where the operator $e^{-\Delta \tau H_{i}}$ acts on, as the system (which includes 4 sites on a plaquette for the ring-exchange interaction model), and the rest of the lattice as the environment.
During the local update process, the environment tensors are fixed, while the system tensors are updated to obtain a best approximation of $\||\Psi_{n+1}\rangle-e^{-\Delta \tau H_{i}}|\Psi_n\rangle\|$. Generally we have to make some approximations to the environment to simplify the calculations.
The quality of the approximated environment often plays a critical role in determining the system tensors during the ITE~\cite{lubasch14_2}.

\begin{figure}[tp!]
	\centering
	\includegraphics[scale=0.4]{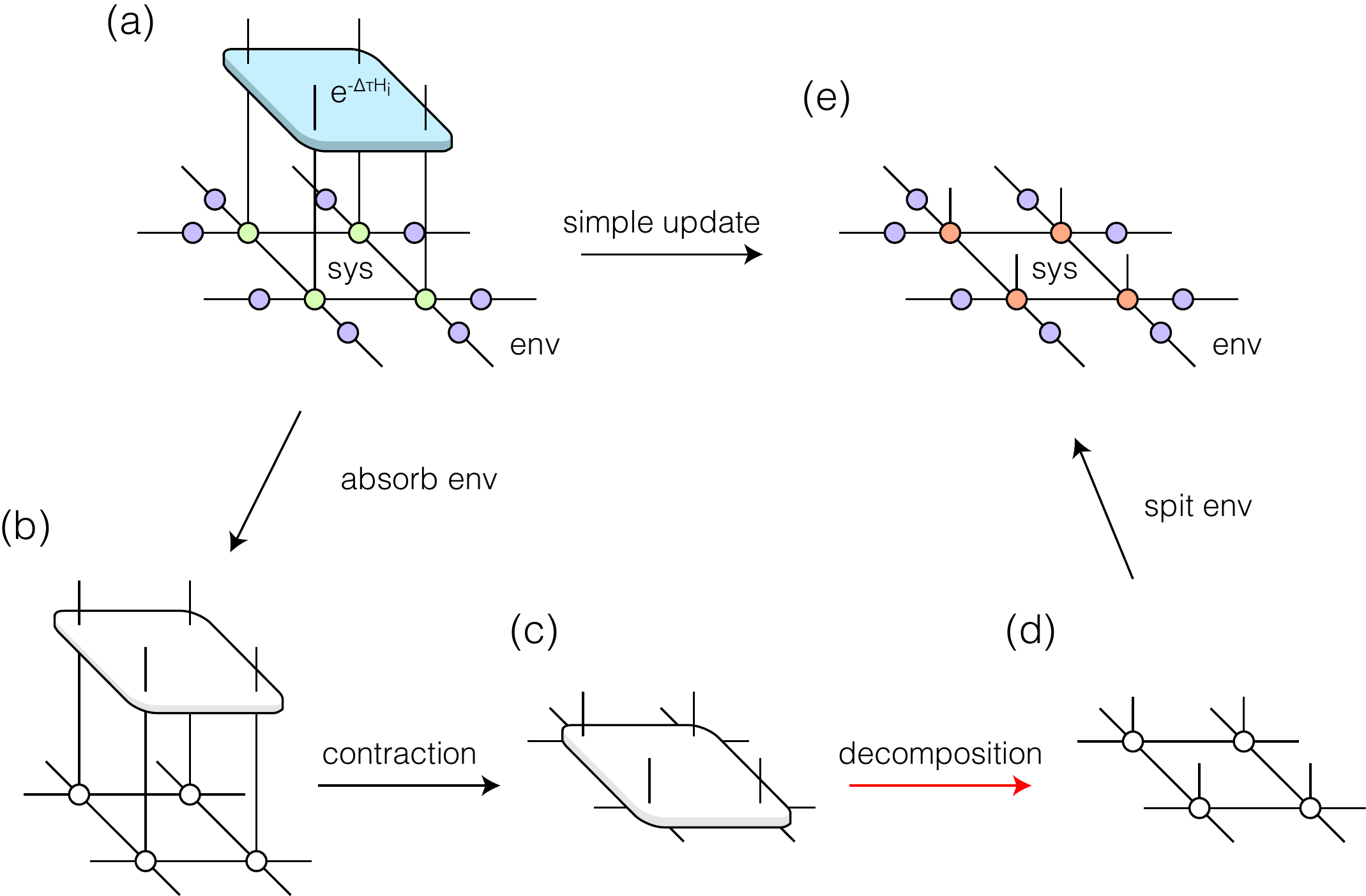}
	\caption{A schematic illustration of a simple update method for the time evolution $|\Psi_{n+1}\rangle =e^{-\Delta \tau H_i}|\Psi_n\rangle$.}
	\label{fig:simple_update}
\end{figure}

For the bond interaction models, a simple update (SU) scheme has been developed \cite{Jiang08} for the local update process,
in which the environment tensor for each virtual bond of the system is assumed to be irrelevant and can be approximated by a real diagonal matrix. Generally, the SU scheme involves four steps, as illustrated in Fig.~\ref{fig:simple_update}:
i) to absorb environment tensors into the system [Fig.\ref{fig:simple_update}(b)]; ii) contract the system tensors with the time evolution operator $e^{-\Delta \tau H_{i}}$ [Fig.\ref{fig:simple_update}(c)]; iii) the contracted system tensor is approximated by a new PEPS [Fig.\ref{fig:simple_update}(d)]; and iv) the environment tensors are split from the new system tensors [Fig.\ref{fig:simple_update}(d)].

However, we can not naively apply the SU to the ring-exchange models for two reasons. Firstly, the third step in the above algorithm is the crucial part of the SU. For the bond interaction models, which only involve two-site interactions, this can be done by the singular value decomposition (SVD) \cite{Jiang08} or the higher-order singular value decomposition (HOSVD)\cite{Xie14}. However, for a ring-exchange interaction $H_{\square,i}$,
there is no simple technique to decompose the tensor in Fig.~\ref{fig:simple_update}(c) into a ring tensor network
in Fig.~\ref{fig:simple_update}(d). Secondly, the SU oversimplifies the environment, which cause serious accuracy and numerical stability problems for the ring-exchange models.

\begin{figure}[tp!]
	\centering
	\includegraphics[scale=0.27]{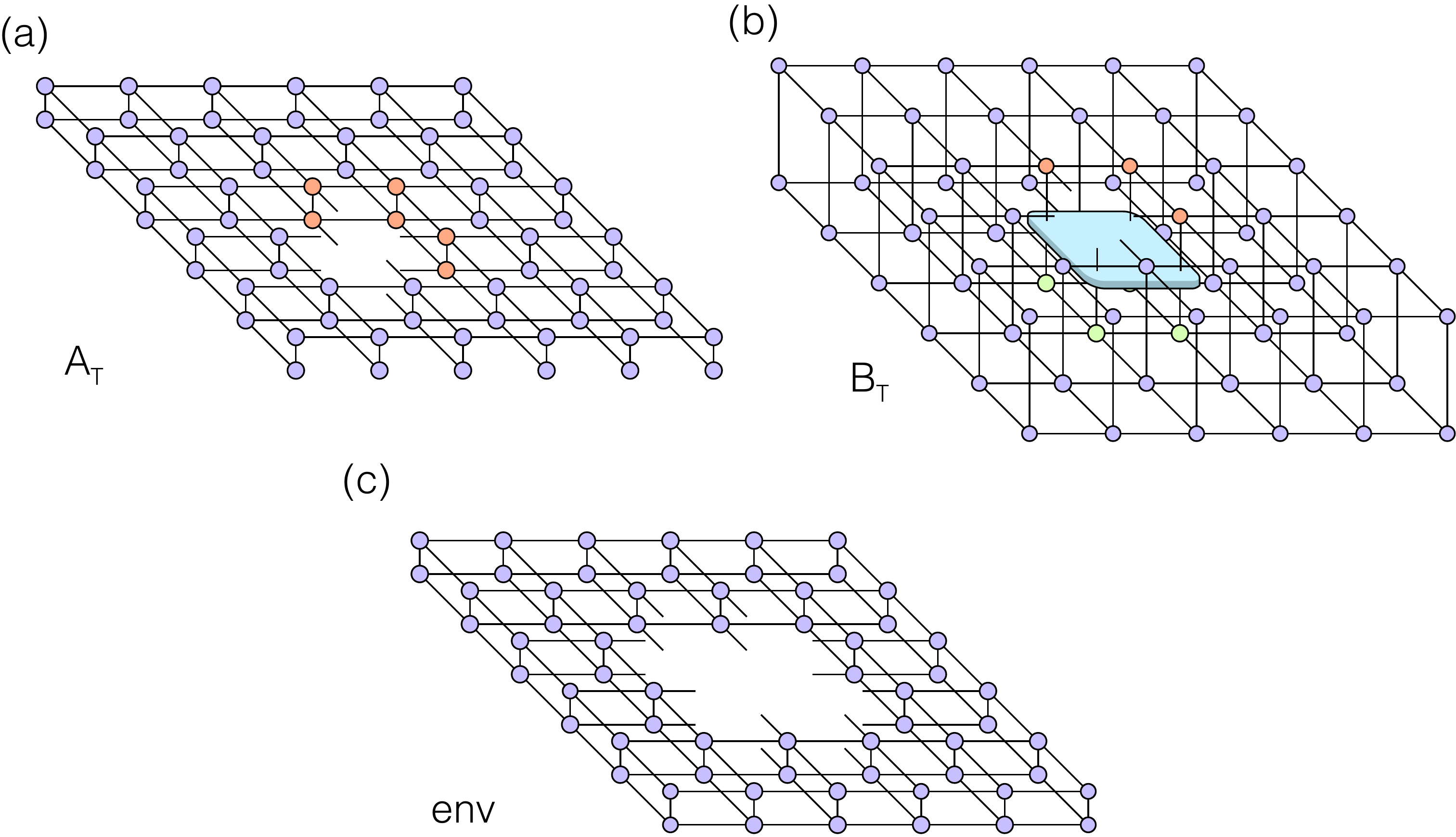}
	\caption{The tensor networks of (a) $A_T$ and (b) $B_T$ for the system tensor $T$ defined in Eq.\ref{eq:AT}
and Eq.\ref{eq:BT}, respectively, and (c) the environment tensor networks for $A_T$ and (b) $B_T$.}
	\label{fig:at_bt}
\end{figure}

Therefore, we have to use an alternative way to perform the local update process by direct minimizing the error function~\cite{lubasch14_2,Haghshenas19},
\begin{equation}
	{\rm Err}(|\Psi_{n+1}\rangle)=\| |\Psi_{n+1}\rangle - e^{-\Delta\tau H_i}|\Psi_n\rangle \| \, ,
	\label{eqn:err_fun}
\end{equation}
where $|\Psi_{n+1}\rangle$ is the new PEPS to approximate
the TNS after time-evolution $e^{-\Delta\tau H_i}$, and $\|\cdot\|$ is the 2-norm of a state.
The error function ${\rm Err}$ is a quadratic function for each tensor in the system,
which can be minimized by iteratively solving the equation
$\partial_{T^*} {\rm Err}(|\Psi_{n+1}\rangle)=0$, i.e.,
\begin{equation}
T\cdot A_T=B_T \, ,
\label{eqn:sys_ten2}
\end{equation}
where,
\begin{eqnarray}
A_T &=& \partial_{T}\partial_{T^*}\langle\Psi_{n+1}|\Psi_{n+1}\rangle,
\label{eq:AT}\\
B_T &=& \partial_{T^*}\langle\Psi_{n+1}|e^{-\Delta\tau H_i}|\Psi_n\rangle
\label{eq:BT}
\, ,
\end{eqnarray}
for each tensor $T$ of $|\Psi_{n+1}\rangle$ in the system part.
The tensor networks that correspond to $A_T$, $B_T$ are shown
in Fig.~\ref{fig:at_bt}(a)-(b).

\subsection{\label{sec:env}Calculation of $A_T$ and $B_T$: approximation of the environment}

To calculate tensor $A_T$ and $B_T$, one has to contract the environment
shown in Fig.~\ref{fig:at_bt}(c).
The environment tensor should be Hermitian and positive semi-definite to ensure the error functions in Eq.\ref{eqn:err_fun} bounded from below\cite{lubasch14_2}. The exact contraction of the environment tensor networks certainly satisfies the conditions, which unfortunately can not be done because the computational cost grows up exponentially with the size of the networks.
The approximations to the environment have been discussed comprehensively in Ref.\cite{lubasch14_2} for bond-interaction models. Here, we discuss the approximation of the environment for the ring-exchange models.

\subsubsection{\label{sect:simple_ene} Simple Environment}

\begin{figure}[tp!]
	\centering
	\includegraphics[scale=0.38]{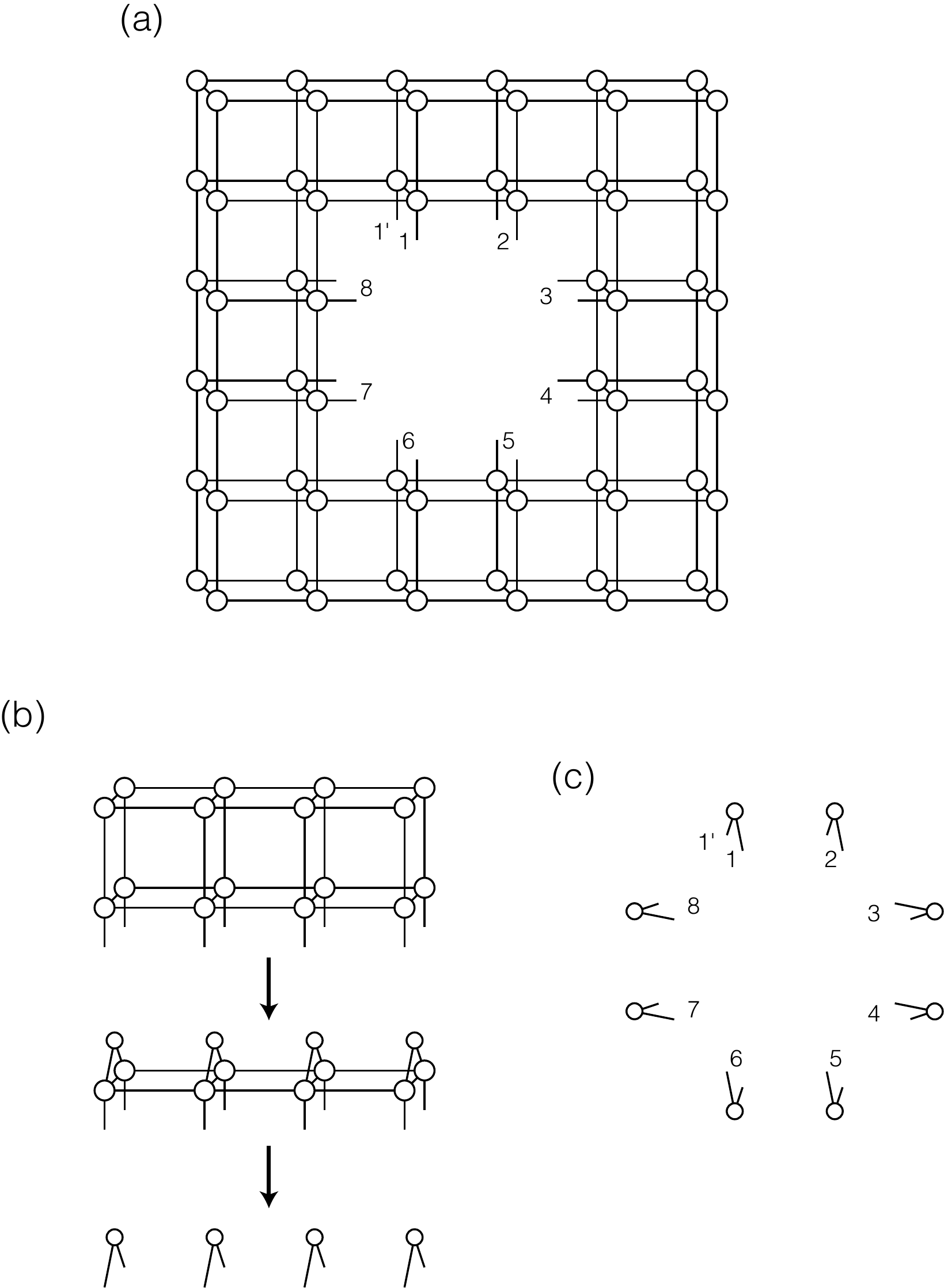}
	\caption{(a) The exact environment of a $2\times 2$ ring tensors in a square PEPS. (b) The simple environment approximation during contraction of environment. (c) The final simple environment as a product state of the tensors
on the 8 sites connected to the system.}
	\label{fig:simp_env}
\end{figure}

If we neglect the entanglement between sites 1 - 8 connected to the system tensors in Fig.\ref{fig:simp_env}(a),
the approximated the environment is just the direct products of Hermitian positive semi-definite matrices on each
site in (bar and ket) pairs [see Fig.\ref{fig:simp_env}(c)]. This is the simplest approximation of the environment,
known as the simple environment\cite{lubasch14,Jiang08}. Obviously the simple environment is automatically Hermitian and positive semi-definite according to its structure.

For an open boundary system, we can obtain the simple environment as follows. We first approximate the tensors at the boundary by the direct products of Hermitian positive semi-definite tensors, by setting $D_c$=1 within the row. We then contract the boundary tensors to the next row, and again approximate
the obtained row of tensors as direct products, as shown Fig.~\ref{fig:simp_env}(b). We repeat this procedure until all environment tensors have been contracted, resulting in the simple environment shown in Fig.~\ref{fig:simp_env}(c).

With the simple environment approximation, tensors $A_T$ and $B_T$ can be easily calculated.
The time complexity to compute $A_T$ and $B_T$ and solve Eq.~\ref{eqn:sys_ten2} is $\mathcal O(D^6)$, compared to
$\mathcal O(D^5)$ for bond interaction models, where $D$ is the bond dimension of the PEPS. This computation cost is relatively low.
The update scheme with the simple environment approximation is also called a generalized SU method
in Ref.~\cite{lubasch14}. The (generalized) SU methods works quite well for many bond interactions models,
however, it may fail for more complicated
ring-exchange models, due to the over simplification of the environment,
which will be discussed in Sec.\ref{sec:mod2}.

\subsubsection{\label{sec:env2}Cluster Environment}

\begin{figure}[htp!]
	\centering
	\includegraphics[scale=0.38]{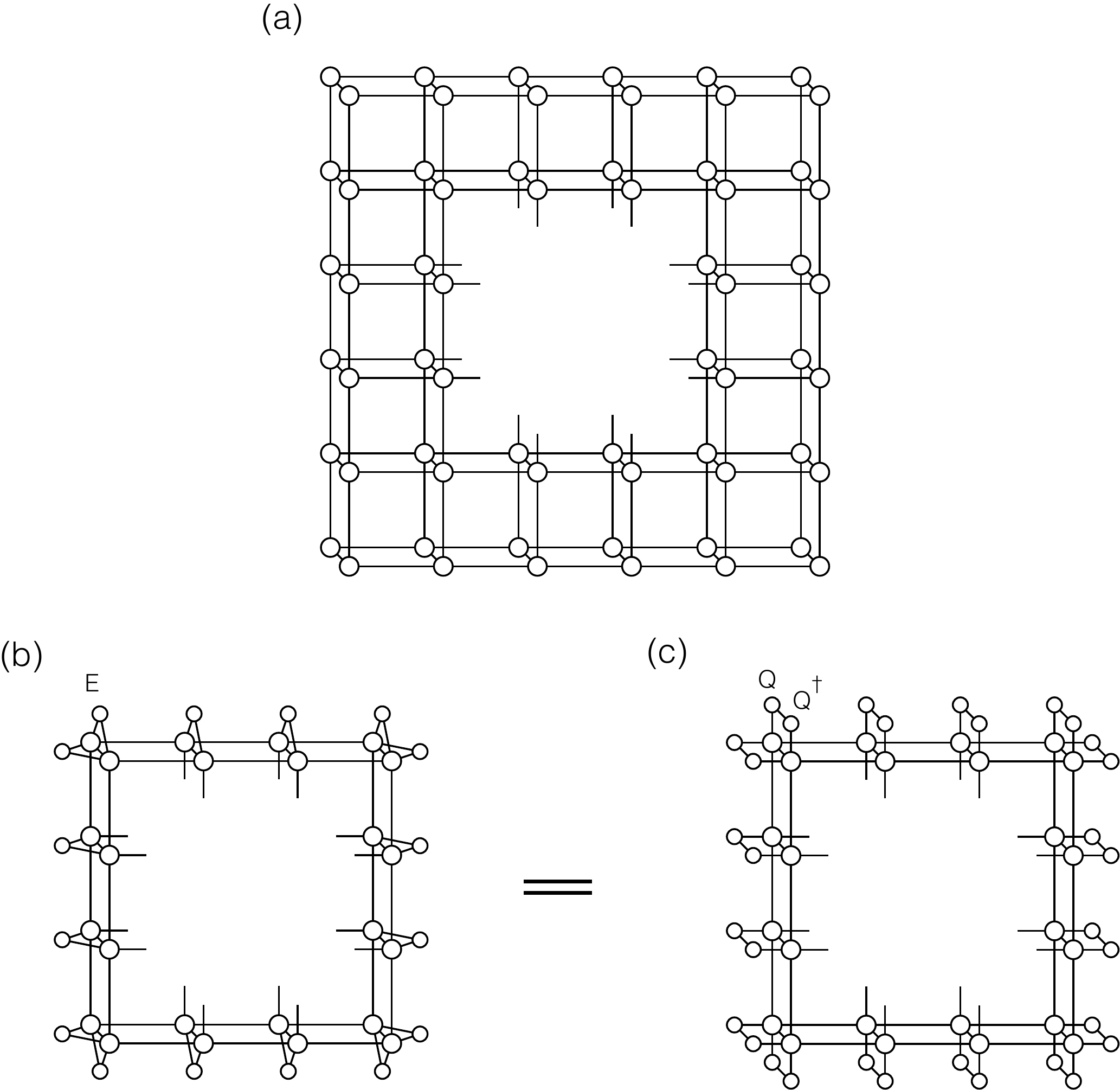}
	\caption{(a) The exact environment of a $2\times 2$ ring tensors in a square PEPS. (b) The cluster environment. (c) The equivalent cluster environment that is obviously Hermitian and positive semi-definite.}
	\label{fig:clt_env}
\end{figure}

To improve the accuracy of the ITE, we need a better approximation to the environment
than the simple environment.
One can directly contract the environment via the  boundary-MPO (bMPO) methods\cite{lubasch14_2,Phien15}. However, if a truncation to the bond dimension, $D_c$ is used, the obtained environment is not guaranteed
to be positive semi-definite. One way to restore the positive semi-definiteness of the environment is to diagonalize the environment tensors and remove all negative eigenvalues~\cite{lubasch14_2}. This procedure works fine for bond interaction models. However, for the ring-exchange model, the final environment tensor is a $D^8\times D^8$ matrix, which is extremely  expensive to diagonalize for large $D$.

Instead, we adopt the cluster environment approximation\cite{lubasch14}. We contract the tensors via boundary MPO method with $D_c$=1 as in the simple environment approximation, except for the tensors which are directly connected with the systems tensors [see Fig.\ref{fig:clt_env}].
To see that the cluster environment is positive semi-definite, we may decompose each simple environment tensor $E$, which is Hermitian and positive semi-definite by construction, into $Q Q^\dagger$, as illustrated by Fig.\ref{fig:clt_env}(b)-(c). By doing so, the cluster environment becomes the contraction of a tensor network with its own conjugate,
which is guaranteed to be Hermitian and positive semi-definite. With cluster environment approximation, the time complexity to compute $A_T$ and $B_T$ and solve Eq.\ref{eqn:sys_ten2} is $\mathcal O(D^{12})$, where $D$ is the bond dimension of original PEPS wave function, which is higher than the simple environment  approximation. The time complexity is also higher than that of cluster update for bond interaction, which is $\mathcal O(D^{10})$~\cite{lubasch14_2}.

\subsection{\label{sec:sing} Reduce the singularity of a PEPS}

After we obtain the (simple or cluster) environment, we may apply time-evolution operators
with ring exchange terms.
However, we find that during the ITE,
there is a strong numerical instability after some time steps.
In previous works\cite{lubasch14_2,Phien15_2,Phien15}, this numerical instability was attributed to the unfixed gauge freedom of PEPS. Some gauge-fixing methods are proposed to avoid the problem. However the relation between the gauge freedom and the numerical instability remains unclear, and there was no standard to choose the best gauge.

To understand the numerical instability during ITE, we define the ``singularity'' of a PEPS,
and we show that the numerical instability is due to the rapid growth of the ``singularity'' of the PEPS.

Let $|\Psi\rangle$ be the PEPS defined in Eq.~\ref{eq:TNS}, and
the singularity of $|\Psi\rangle$ is defined as:
	\begin{equation}
		\sing(|\Psi\rangle)=\frac{\prod_i\| T_i\|}{\langle \Psi | \Psi \rangle} \, ,
	\end{equation}
where $T_i$ is the $i$-th tensor of the tensor network and $\|\cdot\|$ is the 2-norm of a tensor. The singularity of $\Psi$ characterize the  reciprocal of the norm of $\langle \Psi | \Psi \rangle$, when each tensor $T_i$ is normalized to $\| T_i\|$=1. When the norm of a tensor network approaches to 0, the singularity diverges.
In practice, we find that the singularity of the tensor networks grows rapidly during the ITE if untreated,
and the update process becomes numerically unstable~\cite{zhang20}.

To solve this issue, we note that a PEPS has some gauge freedom, e.g., one may insert a pair of invertible gauge matrices $I=GG^{-1}$ in each virtual bond, and the norm of $\langle \Psi | \Psi \rangle$ is invariant.
However, by absorbing the $G$ and $G^{-1}$ matrices to the tensors, $\| T_i\|$ may change, and as a consequence the singularity of the PEPS may vary. Therefore it is possible to reduce the singularity by choosing proper gauges of the PEPS.

Here, we propose a regularization procedure to reduce the singularity of a PEPS. For each bond of the PEPS, we first contract the two tensors connected by the bond, resulting a tensor $C$, and we then make a SVD, $C$=$U\Sigma V$. The tensors of the two sites are updated as $U\Sigma^{1/2}$ and $\Sigma^{1/2}V$, respectively.
This process is illustrated in Fig.\ref{fig:reduce_sig}. As proved in Theorem~\ref{thm:red_sing} in the appendix, this  regularization operations minimize the singularity of the bond tensors, when all other tensors remain fixed. We sweep all the bonds in the PEPS with this procedure for a few iterations, until the singularity of the tensor network converges.

\begin{figure}[tp!]
	\centering
		\includegraphics[scale=0.5]{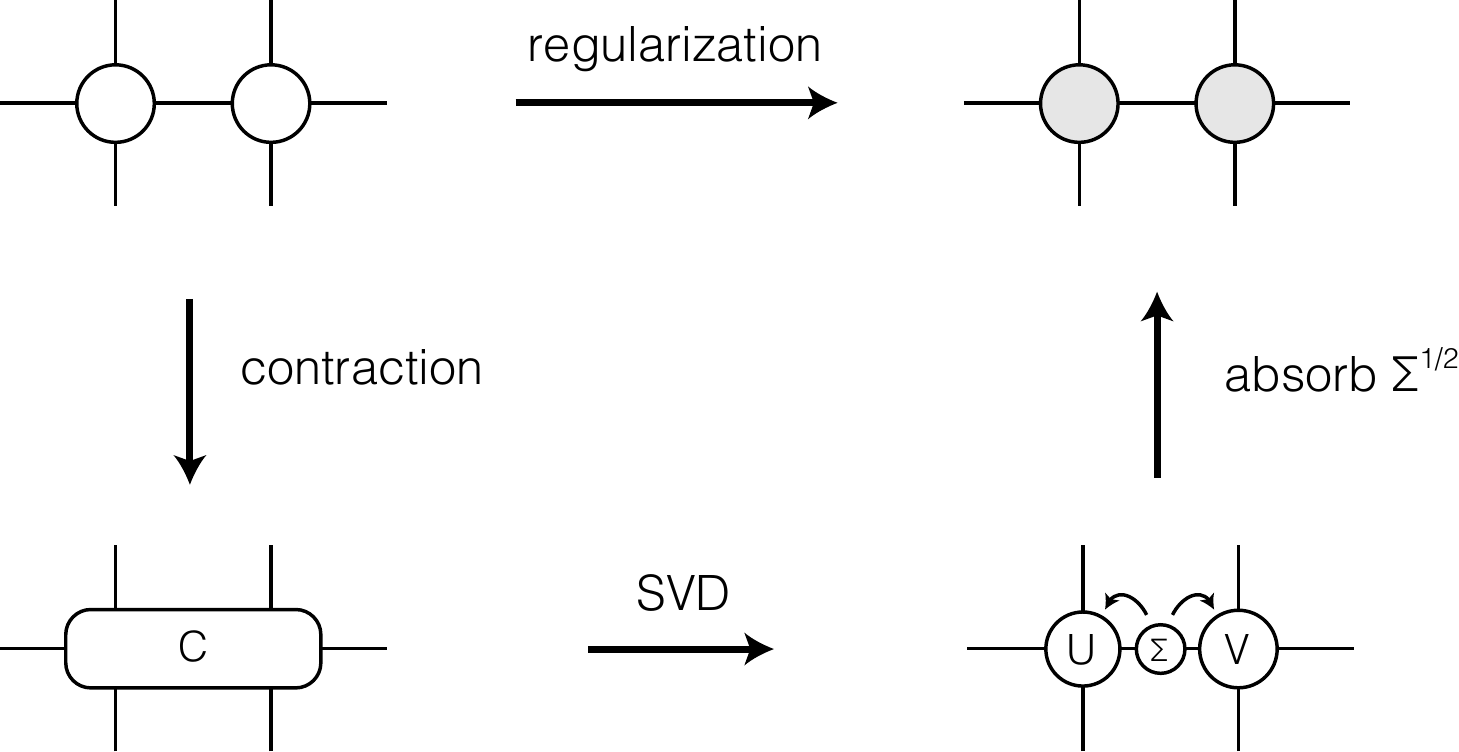}
	\caption{The regularization procedure for a bond in the tensor network.}
	\label{fig:reduce_sig}
\end{figure}

\begin{figure}[tp!]
	\centering
	\includegraphics[width=0.5\textwidth]{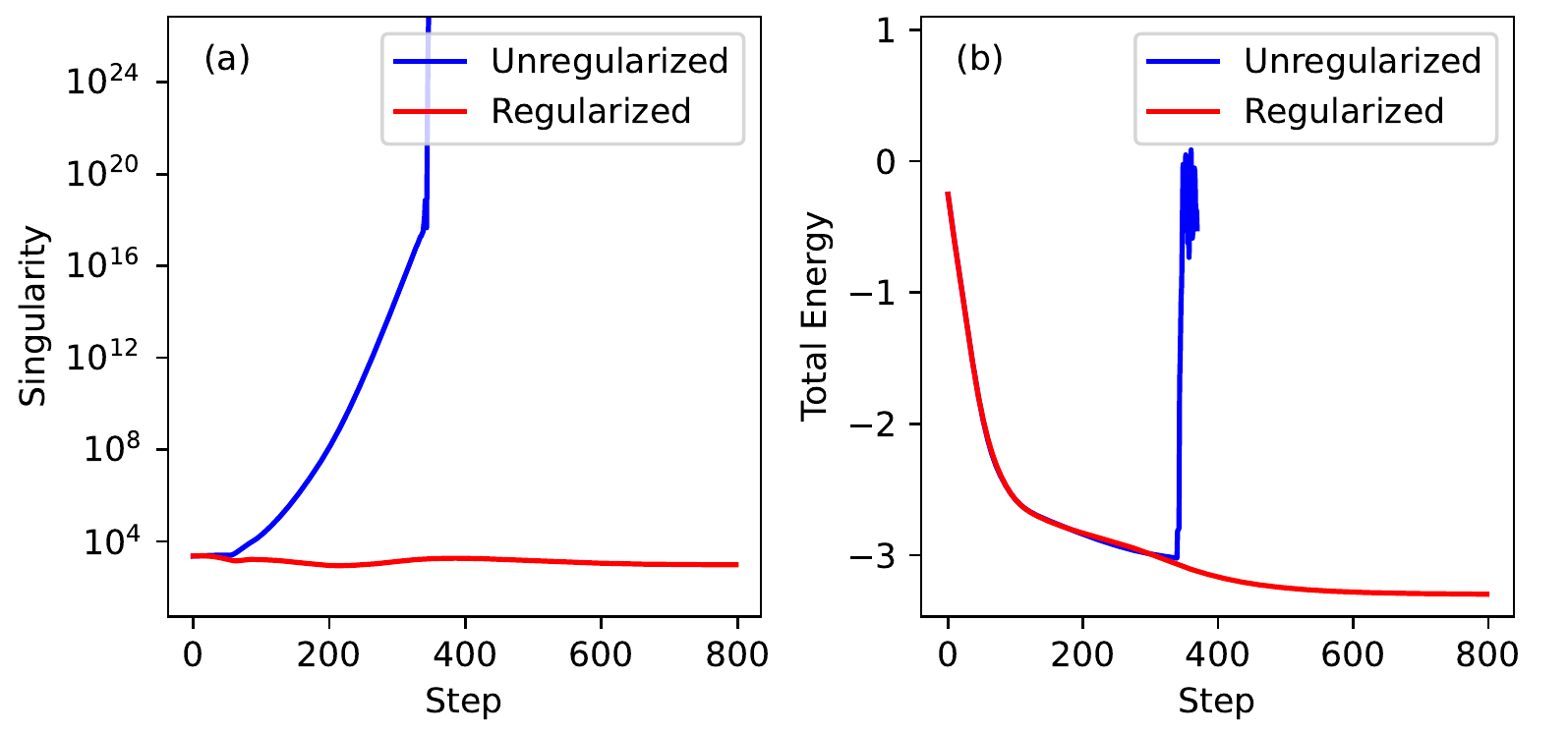}
	\caption{(a) The singularities and  (b) the energies of the PEPS during the ITE optimization
with and without regularization of the singularity.}
	\label{fig:sing_test}
\end{figure}

To demonstrate the effectiveness of the method, we optimize the ground state of a cyclic ring-exchange model introduced in Sec.\ref{sec:mod2}, on a 4$\times$4 square lattice. For convenience, the bond dimension of the PEPS is
taken to be $D$=2. We adopt the cluster environment approximation. During the ITE, we compared the energies and the singularities with and without the regularization process at each time step. The results are shown in Fig.\ref{fig:sing_test}.
As we see, without the regularization procedure, the singularity of the PEPS grows very rapidly with time steps. As a consequence, after 340 steps, the evolution becomes numerically unstable.
In contrast, with the regularization procedure, the singularity remains at low level during the ITE, and the energy can be steadily optimized.

\subsection{Gradient optimization}

Because of the drastic approximation to the environment (even for the cluster environment) during
the ITE, the resulting ground state may not be accurate enough.
It has been demonstrated that the gradient optimization after ITE can significantly improve the
ground state energies for the bond interaction models, where the energy and the gradients are calculated via a Monte Carlo sampling technique with relative low time complexity~\cite{Liu17,He2018,Dong19}. These techniques can also be applied to the ring-exchange models in a similar manner. We perform the gradient optimization after the ITE,
and we show for some ring exchange models, the gradient optimization can significantly improve the ground state energies in the following sections.

\section{\label{sec:mod1}Application to Toric Code Model}

\begin{figure}
	\centering
	\includegraphics[scale=0.45]{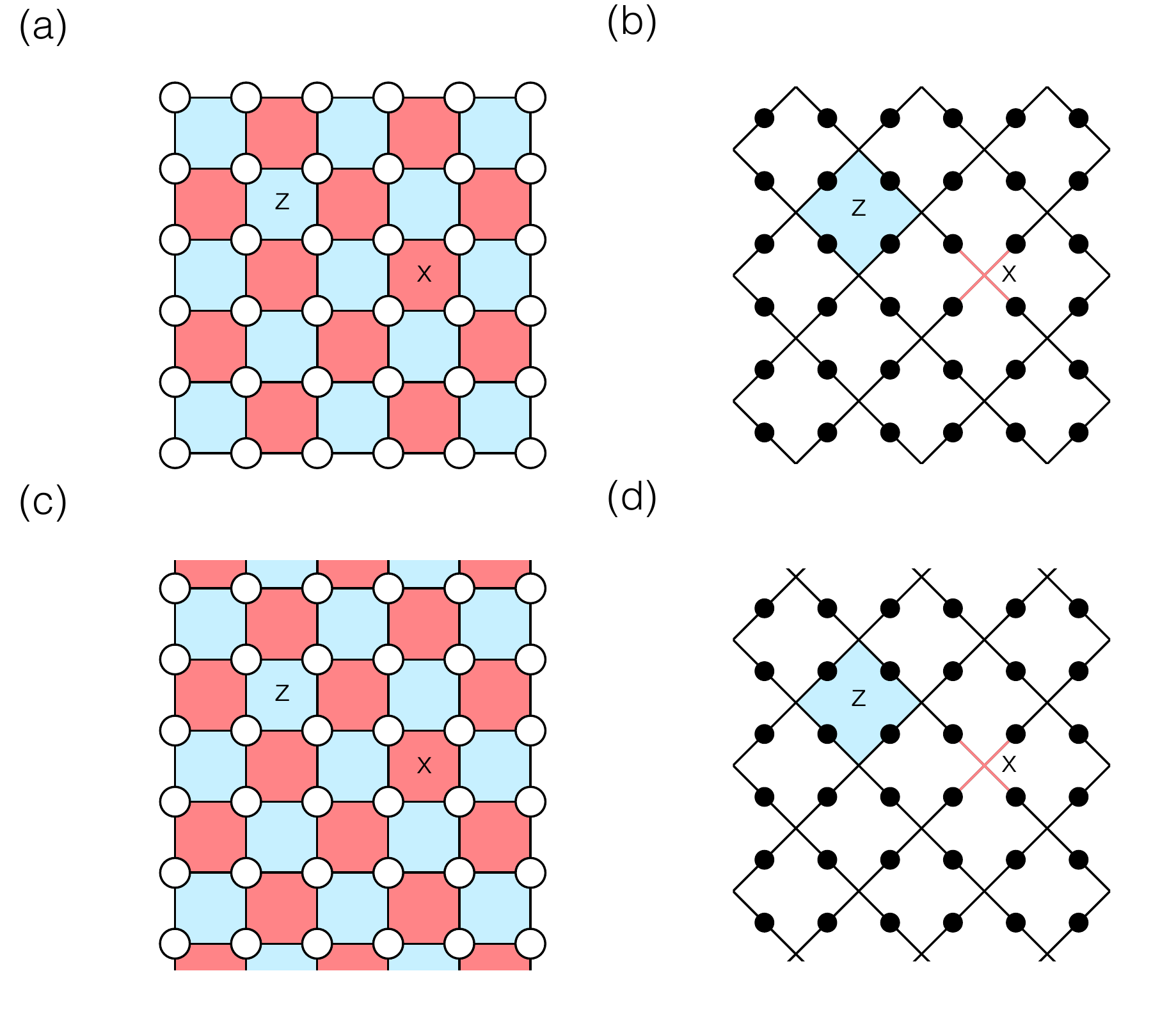}
	\caption{(a)Toric code model on a square lattice with OBC, where spins locate on vertices. $P_z$ are the blue plaquettes and $P_x$ are the red plaquettes. (b) The original Kitaev's toric code model,
where spins locate on edges. (c) and (d) The corresponding toric code models with CBC.}
	\label{fig:toric_code}
\end{figure}

We first apply our method to the toric code model\cite{Kitaev03}.
The toric code model is a paradigm model with ring exchange interactions that can be exactly solved on square lattices.
Generally, the Hamiltonian of the toric code model is defined as,
\begin{equation}
	H=-\sum_{p\in P_x}\prod_{i\in p}\sigma_i^x-\sum_{p'\in P_z}\prod_{i\in p'}\sigma_i^z,
\end{equation}
where $P_x$ and $P_z$ are alternative plaquettes of the lattice as shown in Fig.~\ref{fig:toric_code}, in which
Fig.~\ref{fig:toric_code}(a),(c) corresponding to the OBC and CBC respectively.
We have included appropriate boundary terms, such that our model is equivalent to the Kitaev's original model with smooth boundary conditions, as shown in Fig. \ref{fig:toric_code}(b)(d).

The physical properties, e.g., the topological degeneracy, of the toric code model are closely related to the boundary conditions. The ground state of the toric code model with OBC is non-degenerate, and has two-fold degeneracy on a cylinder\cite{Zhang12}.
More specifically, the ground state of the toric code model on a cylinder has a  gapped
$\mathbb Z_2$ topological order, which has non-trivial low-energy excitations~\cite{Kitaev03}.
The topological entanglement entropy (TEE) in the ground state manifold (GSM),
which is used to character the topological order, is therefore $\log 2$ on a cylinder\cite{Zhang12}.

It has been shown that the ground state of toric code model can be exactly expressed by PEPS with bond dimension $D=2$\cite{gonzales12}. Therefore toric code model serves as a good benchmark for our method. In this section, we benchmark our method by calculating the ground state wave functions of the toric code model on square lattices
and investigating its topological properties.

%\subsection{Ground State Energy}

We first obtain the ground state wave functions of the toric model using PEPS of bond dimension $D=2$, on square lattices of various sizes. Both OBC and CBC are studied, with the smooth boundary conditions \cite{Kitaev98}. We use ITE with simple environment approximation to optimize the ground state, and
the imaginary time step is set to $\Delta \tau$=0.02.
The ground state energies converge very fast with the time steps. The exact ground state energies and relative errors of our numerical results to the exact results of various lattice sizes are listed in Table \ref{tab:toric_ene} for both OBC and CBC. We see that our algorithm can obtain the ground state energies with nearly machine accuracy for the toric code model.

\begin{table}[htb!]
	\centering	
\caption{The exact ground energies of the toric code model of various sizes with both OBC and CBC. The relative errors of the numerical results are less than 10$^{-14}$. }
	\begin{tabular}{cccc}
	\hline\hline
 	Boundary&System&Theoretical &Relative\\
 	 Condition&Size&Energy&Simulation Error\\ \hline
 	\multirow{4}{*}{OBC}&$4\times 4$&-17&\multirow{8}{*}{$<10^{-14}$}\\ %\cline{2-3}
 	&$6\times 6$&-37&\\  %\cline{2-3}
 	&$8\times 8$&-65&\\  %\cline{2-3}
 	&$10\times 10$&-101&\\ \cline{1-3}
 	\multirow{4}{*}{CBC}&$4\times 4$&-16&\\ %\cline{2-3}
 	&$6\times 6$&-36&\\  %\cline{2-3}
 	&$8\times 8$&-64&\\  %\cline{2-3}
 	&$10\times 10$&-100&\\ \hline\hline
 	\end{tabular}
 	\label{tab:toric_ene}
\end{table}

%\subsection{Ground State Renyi Entropy}

\begin{figure}
	\centering
	\includegraphics[scale=0.43]{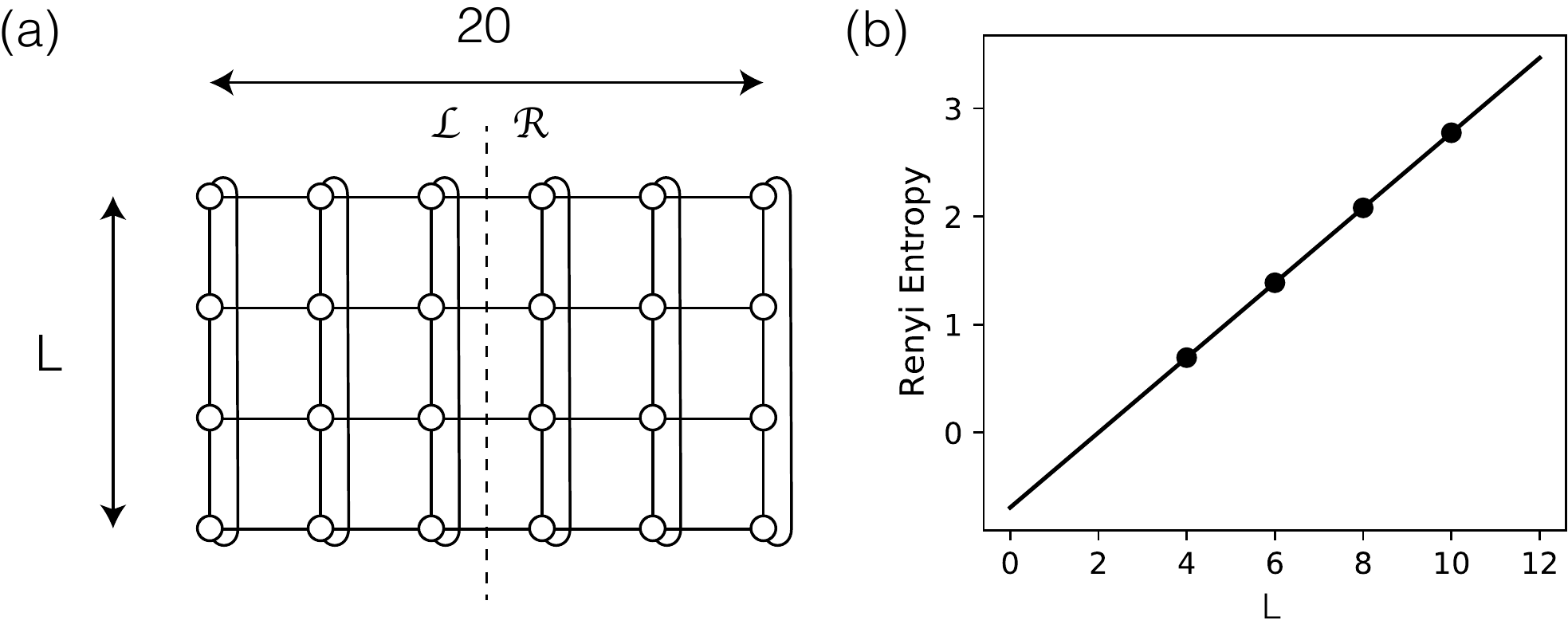}
	\caption{(a) The $L\times 20$ cylindrical lattice, as well as the PEPS, are divided into $\mathcal L$, $\mathcal R$ two parts. (b) Linear fit of $H_2$-$L$ data for toric code model with CBC, where the topological entanglement entropy is $\log 2.003$.}
	\label{fig:division}
\end{figure}

We then calculate the TEE\cite{Kitaev06} of the ground states of the toric code model with CBC
to characterize their topological order.
We divide the cylinder into left ($\mathcal L$) and right ($\mathcal R$) parts, as shown in Fig.\ref{fig:division}.
Let $|s_{\mathcal R}\rangle$ be the basis of the Hilbert space of the part $\mathcal R$, the reduced density matrix of $|\Psi\rangle$ of part $\mathcal L$ is $\rho_{\mathcal L}=\sum_{s_{\mathcal R}}\langle s_{\mathcal R}|\Psi\rangle\langle \Psi|s_{\mathcal R}\rangle$.
The $\alpha$-th order R\'{e}nyi entropy of a state $|\Psi\rangle$
between the two parts is defined as,
\begin{equation}
	H_\alpha=\frac1{1-\alpha}\log \tr(\rho^\alpha_{\mathcal L})\, .
	\label{eqn:renyi_ent}
\end{equation}
Generally, R\'{e}nyi entropy is proportional to the size of the boundary ($L$) between the two parts, i.e.,  when $L \rightarrow\infty$, it has form of,
\begin{equation}
	H_\alpha=-\gamma+a\cdot L+\cdots \, , \label{eqn:tee}
\end{equation}
where the ellipsis represents terms that vanish in the limit of $L\rightarrow\infty$. The constant $\gamma$, which is independent of the size of the system, is the TEE\cite{Flammia09}.
Note that the R\'{e}nyi entropy is ambiguous in a degenerate GSM, and the TEE should be calculated on its minimal entanglement state (MES)\cite{Zhang12}.

To obtain the TEE, we study toric code model on the $L\times$20 rectangular lattices with CBC, which is periodic in the first dimension.
The lattices are divided into the $\mathcal L$ and $\mathcal R$ parts, both have the size of $L\times$10 as shown in Fig.\ref{fig:division}(a). We calculate the 2nd order R\'{e}nyi entropy between the two parts of the ground states in the whole GSM. We then obtain the minimal R\'{e}nyi entropy in the GSM. More details can be found in Appendix~\ref{sec:ent}.

The numerically calculated R\'{e}nyi entropies of the MES are listed in Table~\ref{tab:toric_ent}, for $L$=4 - 10, which are in excellent agreement with exact results. The R\'{e}nyi entropy shows excellent linearity with $L$, as shown in Fig.~\ref{fig:division}(b), and the fitted TEE for the toric code model with CBC is $\log 2.003$, which is in excellent agreement the theoretical result $\log 2$.

\begin{table}[htb!]
	\centering	
\caption{The minimal R\'{e}nyi entropy of the toric code model on $L\times$20 cylinders.}
	\begin{tabular}{cccc}
	\hline\hline
	System&Boundary&Theoretical&Simulation Value\\
 	Size&Length  &$H_2$ for MES&of $H_2$ for MES\\ \hline
 	$4\times 20$&4&$\log 2$&$\log(2+2.8\times 10^{-11})$\\
 	$6\times 20$&6&$\log 4$&$\log(4-1.5\times 10^{-5})$\\
 	$8\times 20$&8&$\log 8$&$\log(8+1.6\times 10^{-5})$\\
 	$10\times 20$&10&$\log 16$&$\log 16.032$\\
 \hline\hline
 	\end{tabular}
 	\label{tab:toric_ent}
\end{table}

\section{\label{sec:mod2}Application to cyclic Ring Exchange Model}

In this section,  we investigate the ground state of the cyclic ring exchange model \cite{Chubukov92,Kolezhuk98,Brehmer99,Muller02,Lauchli03,Gritsev04,Delannoy05,Song06,Delannoy09}
using the PEPS method developed in this work.
The cyclic ring exchange model on a square lattice can be written as,
\begin{align}
	H
	=&\sum_{(ijkl)}\big[(S_i\cdot S_j+\frac 14)(S_k\cdot S_l+\frac 14)+(S_i\cdot S_l+\frac 14)\nonumber\\
	&(S_k\cdot S_j+\frac 14)-(S_i\cdot S_k-\frac 14)(S_j\cdot S_l-\frac 14)-\frac 1{16}\big]\nonumber\\
	=&\sum_{(ijkl)}\big[\frac 14 P_{ijkl}+h.c. -\frac 1{16}\big] \, ,
\label{eqn:hami}
\end{align}
where $(ijkl)$ denotes a $2\times 2$ plaquette with four sites $i,j,k,l$ in the clockwise order. Here, $i$=$(i_x, i_y)$ are the site index of the spin.
The operator $P_{(ijkl)}$ performs a cyclic permutation of spins on $i,j,k,l$ sites.
The cyclic ring exchange interactions arise naturally as higher-order terms in the magnetic effective model
of half-filled Hubbard model in the strong interaction limit\cite{MacDonald88},
which plays an important role in many materials \cite{Katanin02,Calzado03,Toader05,Delannoy09}.

The cyclic ring exchange model on a quasi-1D ladder lattice with width $L= 4$ has been investigated using the DMRG method~\cite{Lauchli03}. The results indicate that the ground state of the model has a strong vector chiral order.
Here we investigate the cyclic exchange model on the genuine 2D lattices.

%\subsection{Simulation of the Ground State}

We calculate the ground states of the cyclic ring exchange model on a $L\times L$ square lattice with $L$=4 - 12 via PEPS. The maximum bond dimension $D$= 8 is used. During the calculations, we first perform ITE optimization using the cluster environment.
The simulation is done with imaginary time step $\Delta \tau=0.02$ and is proceeded until the energy (computed by MC sampling method \cite{Liu17}) is converged.  We start from bond dimension $D$=2, and gradually increase $D$. We use the optimized PEPS of $D$ as the starting PEPS of $D$+1.
To obtain the highly accurate ground state, we further optimize the PEPS via the gradient optimization method~\cite{Liu17} after ITE for each bond dimension $D$.

\begin{figure}
	\centering
	\includegraphics[scale=0.58]{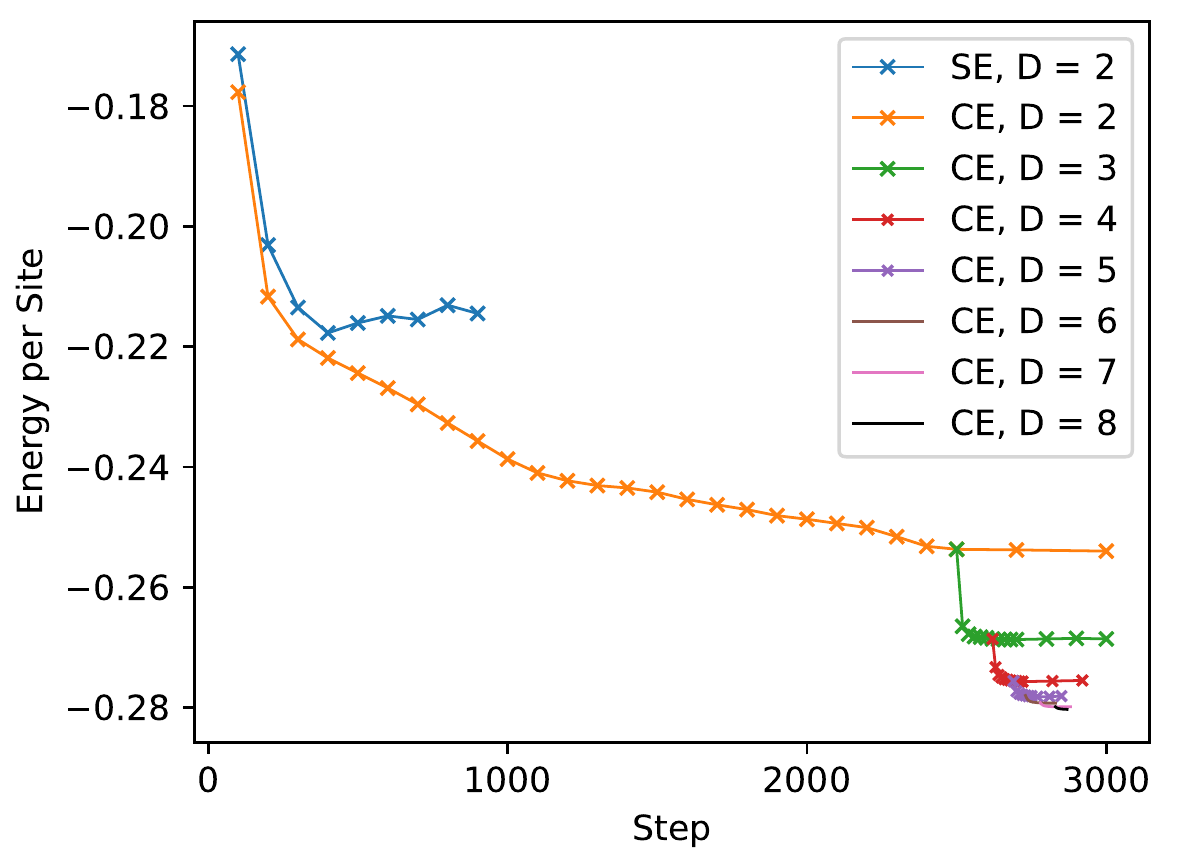}
	\caption{The ground state energy of the cyclic ring-exchange model on the $10\times 10$ lattice during ITE,  with simple (SE) and cluster environment(CE). The bond dimensions are taken as $D$= 2 - 8. }
	\label{fig:ringene1}
\end{figure}
\begin{figure}
	\centering
	\includegraphics[scale=0.58]{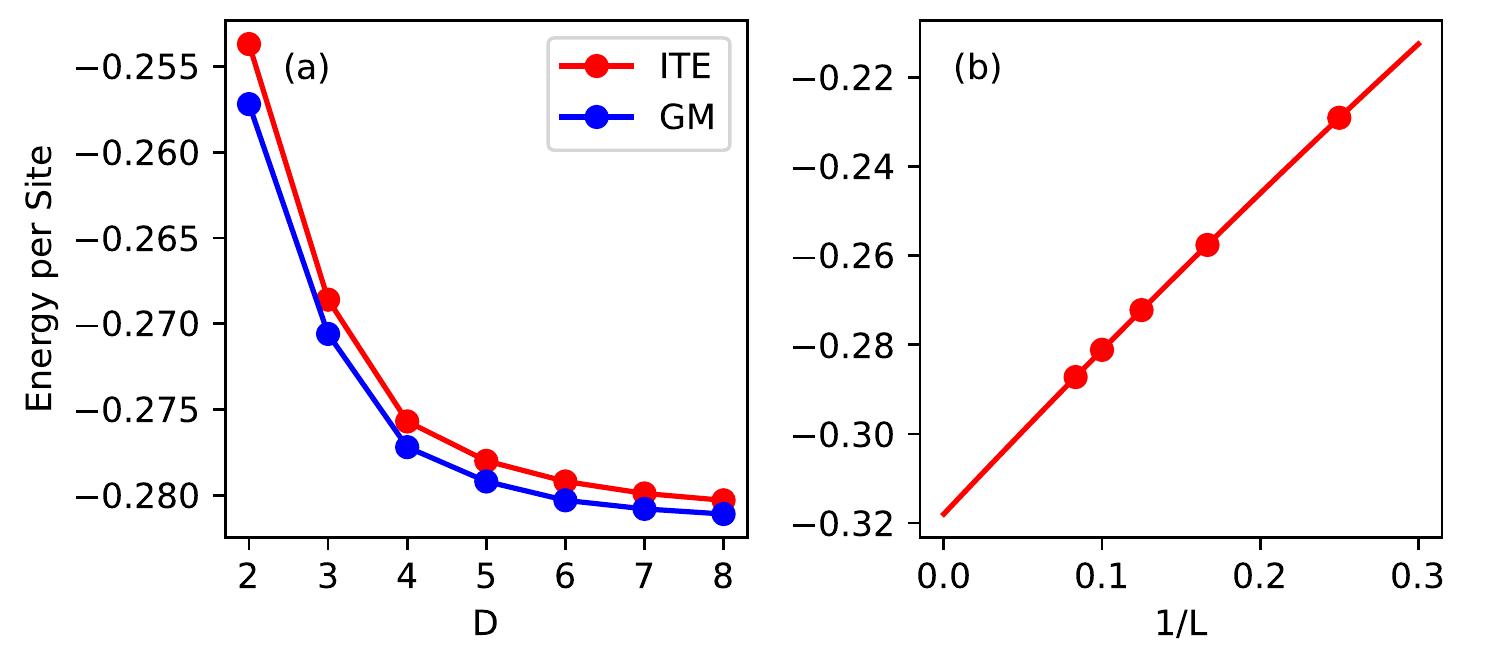}
	\caption{(a)Comparing the ground state energy of the cyclic ring-exchange model on the $10\times 10$ lattice, obtained by the gradient optimization (GM) to ITE for bond dimensions $D$=2 - 8.
(b) The finite size scaling of the ground state energy obtained by GM with bond dimension $D$=8.}
	\label{fig:ringene2}
\end{figure}

The ground state energies as functions of imaginary time steps on a $10\times 10$ lattice are
shown in Fig. \ref{fig:ringene1} for different bond dimensions $D$.
As a comparison, we also show the results obtained by simple environment for $D$=2.
We see that the energy can be optimized steadily using the cluster environment,
but not with the simple environment.

In Fig.\ref{fig:ringene2}(a), we compare the ground state energies of different bond dimensions obtained by the ITE to those by the gradient optimization method for the 10$\times$10 lattice.
We see that the gradient optimization can significantly improve the ground state energy.
For the 10$\times$10 lattice, the ground state energy of the model converges very well at $D$=8.
To obtain the ground state energy in the thermodynamic limit, we perform finite size scaling using the energies,
obtained by the gradient optimization for bond dimension $D$=8, %of $L$=4, 6, 8, 10, 12,
as shown in Fig.\ref{fig:ringene2}(b).
The ground state energy per site in the thermodynamic limit is $E_\infty$=-0.3180, determined by a second order polynomial fitting.

%\subsection{Order Parameters}

To determine the possible ordering in the ground state, we calculate spin-spin correlation function
$S_{ij}=\langle S_i \cdot S_j\rangle-\langle S_i \rangle\cdot\langle S_j\rangle$,
and dimer-dimer correlation function $D^\kappa_{ij}=\langle D^\kappa_i D^\kappa_j\rangle-\langle D^\kappa_i\rangle\langle D^\kappa_j\rangle$ ($\kappa=x/y$) where $ D^{x/y}_i=S_i \cdot S_{i+e_x/e_y}$.
Their Fourier transformations,
\begin{eqnarray}
S_{\bf k} &=& \frac{1}{L^4}\sum_{ij}e^{i {\bf k}\cdot(i-j)}S_{ij} \, , \\
D^\kappa_{\bf k}&=& \frac{1}{L^2(L-1)^2}\sum_{ij}e^{i{\bf k}\cdot(i-j)}D^\kappa_{ij}  \, ,
\end{eqnarray}
can be used as the order parameters.
Especially, $S_{(\pi,\pi)}$, $S_{(0,\pi)}$ are the order parameters of the N\'{e}el order and stripe order, respectively; whereas $D^\kappa_{(\pi,\pi)}$, ($\kappa$=$x$, $y$) are the order parameters of the horizontal/vertical dimer orders.

Besides the N\'{e}el, stripe and dimer order, a vector chiral phase that breaks space-inversion symmetry is also possible \cite{Lauchli03}. We also calculate the vector chiral-chiral correlation function,
\begin{equation}
V_{mn}=\langle V_m\cdot V_n\rangle-\langle V_m\rangle\cdot\langle V_n\rangle \, ,
\end{equation}
where $V_m=S_i \times S_j+S_j \times S_k+S_k \times S_l+S_l \times S_i$ is a vector operator on plaquette $m$, and $i,j,k,l$ are four vertices of the plaquette in clockwise order. Its Fourier transformation,
\begin{equation}
V_{\bf k}=\frac{1}{(L-1)^4}\sum_{ij}e^{i{\bf k}\cdot(i-j)}V_{ij} \, ,
\end{equation}
at ${\bf k}$=${\bf (\pi,\pi)}$, is the order parameter of the vector chiral order.

\begin{figure}
	\centering
	\includegraphics[scale=0.6]{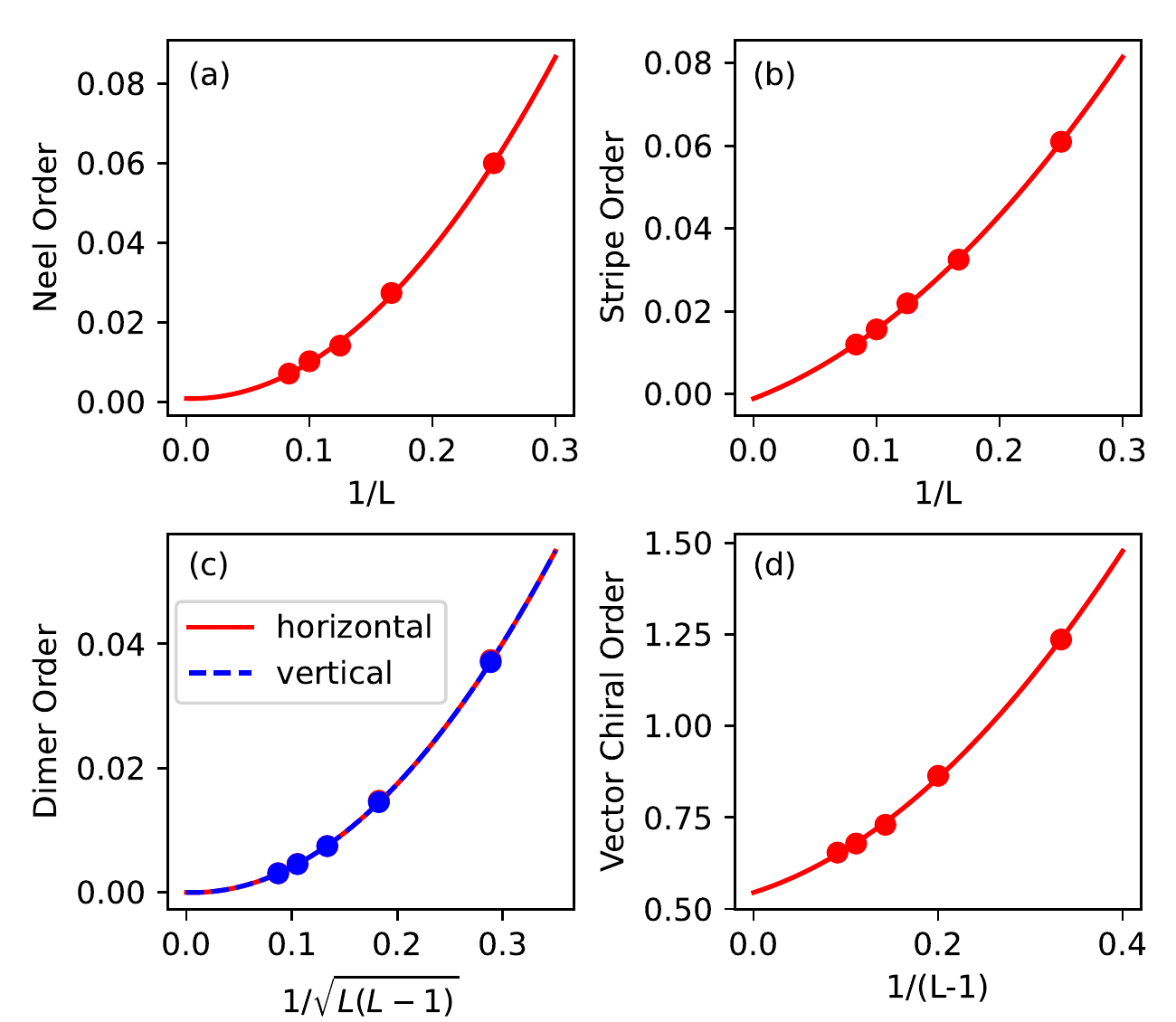}
	\caption{Finite size scaling of the (a) N\'{e}el order parameters $S_{(\pi,\pi)}$; (b) stripe order parameters $S_{(0,\pi)}$; (c) horizontal/vertical dimer order parameters $D^{x/y}_{(\pi,\pi)}$; (d) vector chiral order parameters $V_{(\pi,\pi)}$.}
	\label{fig:order_para}
\end{figure}

The finite size scaling of the order parameters %(at $D$=8)
are shown in Fig.\ref{fig:order_para}.
We find that the N\'{e}el order, stripe order and dimer order all vanish in the thermodynamic limit ($L\rightarrow \infty$), via the second order polynomial fitting. However, the vector chiral order parameter $V_{\pi,\pi}=0.545$ remains large in the thermodynamic limit. These results suggest that the cyclic ring exchange model has strong vector chiral order~\cite{Lauchli03} in the two-dimensional lattice.

\section{\label{sec:con}Conclusion and Outlook}

We generalize the ITE to optimize PEPS wave functions for the ring-exchange models on two-dimensional lattices. We compare the effects of different approximations to the environment. We propose a scheme
to reduce the singularity of the PEPS, which can significantly improve the numeric stability during the ITE. We benchmark our method with the toric code model, and obtain extremely accurate ground state energies and topological entanglement entropy. We also benchmark our method with the two-dimensional cyclic ring exchange model, and find that the ground state has a strong vector chiral order. This method can be a powerful tool to investigate the models with ring interactions, e.g., the Bose metal model \cite{Motrunich07,Sheng08,Block11,Mishmash11,Huerga14} in genuine 2D systems. The methods developed in this work, e.g., the regularization process to reduce the singularity can also be applied to other models.

\acknowledgments

We are grateful to the helpful discussion with C-J Wang.
The work is supported by the National Science Foundation of China (Grant Number 11774327, 11874343).
The numerical calculations have been done on the USTC HPC facilities.

\appendix

\section{\label{sec:sing_tn}Singularity of Tensor Network}

Let ${\rm TN}$  be a tensor network (e.g. a PEPS) that consists of tensors $\{T_i\}$, such that the contraction of  ${\rm TN}$, $\contract({\rm TN} )\neq 0$, the singularity of ${\rm TN}$ is defined as,
	\begin{equation}
		\sing({\rm TN})=\frac{\prod_i\| T_i\|}{ \| \contract({\rm TN} )\|} \, ,
\label{eq:singularity}
	\end{equation}
where $\|\cdot\|$  stands for the 2-norm of a tensor, unless otherwise noticed.
We say that the two tensor networks of the same structure (i.e., they have the same bond connections and bond dimensions) are equivalent if they give the same tensor after contraction. For a given  ${\rm TN}$ ($\contract(\rm TN)\neq 0$), we would like to find the equivalent ${\rm TN}$ with minimal singularity, which we call the minimal singularity form of a ${\rm TN}$.

Let $\mathcal{B}(A,B)=$ \adjustbox{raise=-1.8ex}{\begin{tikzpicture}[tensor/.style={circle,fill=white,draw}]
		\draw 	(-1.6,0) -- (1.6,0);
		\node [tensor] (A) at (-0.7,0) {$A$};
		\node [tensor] (B) at (0.7,0) {$B$};
		\node at (-1.3,0.25) {$M$};
		\node at (0,0.25) {$N$};
		\node at (1.3,0.25) {$K$};
	\end{tikzpicture}}
be a tensor network with bond dimensions $M$, $N$, $K$, respectively. We may treat $A$ and $B$ as matrices, and the contraction of the tensor network is just the matrix product $AB$. For this particularly simple kind of tensor networks, its minimal singularity form exists and can be rigorously obtained, as follows.
\begin{theorem}
The minimal singularity form of $\mathcal{B}(A,B)$ exists.\label{thm:exist}
\end{theorem}
\begin{proof}
	We define the continuous function $\sing_{\mathcal B} (A,B):\mathbb C^{M\times N}\times\mathbb C^{N\times K}\rightarrow \mathbb R$ by
	\begin{equation}
		\sing_{\mathcal B}(A,B)=\sing(\mathcal B(A,B))=\frac{\|A\|\|B\|}{\|AB\|}\, .
	\end{equation}
We need to prove that $\sing_{\mathcal{B}}$ can reach its minimal value in the closed subset $S=\{(A',B')|A'B'=AB\}$.
	
	According to Theorem 4.28 in \cite{Apostol}, $\sing_{\mathcal{B}}$ can reach its minimal value in the compact subspace $S'=\{(A',B')|\|A'\|\leq\|A\|,\|B'\|\leq\|B\|,A'B'=AB\}$. Let the minimum point be $(A_0,B_0) \in S'$.
If $(A_0,B_0)$ is not the minimum point in $S$, i.e., there exists $(A_1,B_1)\in S$, such that $\sing_{\mathcal B}(A_1,B_1)<\sing_{\mathcal B}(A_0,B_0)$, which implies that $\|A_1\|\|B_1\|<\|A_0\|\|B_0\|$. Therefore, we can always
find a proper real number $\lambda>0$, such that $(\lambda A_1,\frac 1\lambda B_1)\in S'$, and $\sing_{\mathcal B}(\lambda A_1,\frac 1\lambda B_1)=\sing_{\mathcal B}(A_1,B_1)<\sing_{\mathcal B}(A_0,B_0)$. This contradicts to the fact that $(A_0,B_0)$ is the minimum point in $S'$. Therefore, $(A_0,B_0)$ must be the minimum point in $S$.
\end{proof}

\begin{lemma}
	Let $A$ be a complex matrix of shape $M\times N$. Let $A^\dagger A=W^\dagger\Sigma W$ be a diagonalization. By appropriate permutation, we can always ensure that $\Sigma=\diag(\lambda_1,\dots,\lambda_n,0,\dots)$, where $\lambda_i>0$. Then we have $A=V_A\Sigma_A W$ for some unitary $V_A$, where $\Sigma_A=\diag(\sqrt\lambda_1,\dots,\sqrt\lambda_n,0,\dots)$ is a diagonal matrix of shape $M\times N$.
	\label{lem:red_sing}
\end{lemma}
\begin{proof}
	Similar to the proof of Theorem 11.4 in Ref.~\cite{Serre10}.
\end{proof}

\begin{theorem}
$\mathcal{B}(A_0,B_0)$, where $A_0B_0=AB$, is the minimal singularity form of $\mathcal{B}(A,B)$ if and only if
\begin{equation}
\label{eqn:bond_sing_suf0}
	A_0^\dagger A_0/\|A_0\|^2 =B_0B_0^\dagger/\|B_0\|^2 \, .
%\label{eqn:bond_sing_suf}
\end{equation}
\label{thm:red_sing}
\end{theorem}

\begin{proof}

	We first prove that Eq.\ref{eqn:bond_sing_suf0} is the necessary condition. Suppose that $\mathcal B(A_0,B_0)$ has minimal singularity. For an arbitrary infinitesimal matrix $\delta S$, let $A'=A_0e^{\delta S}$ and $B'=e^{-\delta S}B_0$, and the singularity of the tensor network $\mathcal B(A',B')$ is not less than that of $\mathcal B(A_0,B_0)$, i.e.,
	\begin{align}
		\| A'\|^2\| B'\|^2=&\| A_0e^{\delta S}\|^2\| e^{-\delta S}B_0\|^2 \nonumber\\
		=&\| A_0(1+\delta S)\|^2\| (1-\delta S)B_0\|^2+\mathcal O(\delta S^2) \nonumber\\
		=&\| A_0\|^2\|B_0\|^2+2{\rm Re}\big[\tr[(A_0^\dagger A_0\|B_0\|^2-\nonumber \\
		&B_0B_0^\dagger\|A_0\|^2)\delta S]\big]+\mathcal O(\delta S^2) \nonumber\\
		\geq& \| A_0\|^2\|B_0\|^2 \, .
	\end{align}
	
	For the above inequality to hold for any $\delta S$, we must have
	\begin{equation}
		A_0^\dagger A_0\|B_0\|^2-B_0B_0^\dagger\|A_0\|^2=0 \, .
	\end{equation}
	It thus proves that Eq.\ref{eqn:bond_sing_suf0} is a necessary condition for Theorem~\ref{thm:red_sing}.

	Next, we prove Eq.\ref{eqn:bond_sing_suf0} is the sufficient condition for Theorem~\ref{thm:red_sing}.
Let $A_0$ and $B_0$ satisfy Eq.\ref{eqn:bond_sing_suf0}, then we have,
	\begin{equation}
		\bigg(\frac{A_0}{\|A_0\|}\bigg)^\dagger\frac{A_0}{\|A_0\|}
      =\frac{B_0}{\|B_0\|}\bigg(\frac{B_0}{\|B_0\|}\bigg)^\dagger
	\end{equation}
	
	By Lemma~\ref{lem:red_sing}, we have,
	\begin{align}
		\frac{A_0}{\|A_0\|}&=V_{A_0}\Sigma_{A_0} W\\
		\bigg(\frac{B_0}{\|B_0\|}\bigg)^\dagger&=V_{B_0}\Sigma_{B_0} W
	\end{align}
	where $\Sigma_{A_0}$ and $\Sigma_{B_0}$ have identical nonzero diagonal elements, which are arranged before zero diagonal elements. Therefore,
\begin{equation}
		A_0B_0 %=V_{A_0}(\|A_0\|\|B_0\|\Sigma_{A_0}\Sigma_{B_0}^T)V_{B_0}^\dagger
=V_{A_0}\Sigma_{A_0B_0} V_{B_0}^\dagger \, ,
\label{eqn:AB_sing}
\end{equation}
where $\Sigma_{A_0B_0}=\|A_0\|\|B_0\|\Sigma_{A_0}\Sigma_{B_0}^T$.

Let $\Sigma_{A_0B_0}=\diag(\lambda_1,\dots,\lambda_n,0,\dots)$ ($\lambda_i>0$). Since the right hand side of Eq.~\ref{eqn:AB_sing} is the SVD of matrix $A_0B_0$, $\{\lambda_i\}$ are the singular values $A_0B_0$. Therefore, $\sum \lambda_i$ , which is also known as the nuclear norm $\| A_0B_0\|_*$, is completely determined by the matrix $A_0B_0=AB$.
Since $\|A_0\|\Sigma_{A_0}$ and $\|B_0\|\Sigma_{B_0}^T$ have proportional nonzero diagonal elements, we have
	\begin{align}
		\|A_0\|\Sigma_{A_0}&=m\diag(\sqrt\lambda_1,\dots,\sqrt\lambda_n,0,\dots)\label{eqn:bond_sing_A}\\
		\|B_0\|\Sigma_{B_0}^T&=\frac 1m\diag(\sqrt\lambda_1,\dots,\sqrt\lambda_n,0,\dots)\label{eqn:bond_sing_B}
	\end{align} for some undetermined real number $m>0$.
The singularity of $\mathcal{B}(A_0,B_0)$ is then,
	\begin{equation}
		\sing(\mathcal{B}(A_0,B_0))=\frac{\|A_0\|\|B_0\|}{\|A_0B_0\|}=\frac{\| AB\|_*}{\|AB\|}\, .
\label{eq:singAB}
	\end{equation} Therefore, for any $(A_0,B_0)$ that satisfies Eq.~\ref{eqn:bond_sing_suf0}, $\mathcal{B}(A_0,B_0)$ has the same singularity.
Since we have proven that a minimal singularity form of $\mathcal{B}(A,B)$ must satisfy Eq.~\ref{eqn:bond_sing_suf0},  it thus proves that Eq.~\ref{eqn:bond_sing_suf0} is also a sufficient condition
for Theorem \ref{thm:red_sing}.

\end{proof}

Particularly, $A_0$ and $B_0$ can be chosen as follows.
Let $U\Sigma V$ be the SVD of the matrix $AB$, where $\Sigma$ is a quasi-diagonal singular matrix $\diag(\lambda_1,\dots,\lambda_n,0,\dots)$ of shape $M\times K$, such that $\lambda_i>0$. Let $\bar\Sigma_A$ be a
quasi-diagonal matrix $\diag(\sqrt\lambda_1,\dots,\sqrt\lambda_n,0,\dots)$ of shape $M\times N$, and $\bar\Sigma_B$ be a quasi-diagonal matrix $\diag(\sqrt\lambda_1,\dots,\sqrt\lambda_n,0,\dots)$ of shape $N\times K$. It is easy to check that $A_0=U\bar\Sigma_A$ and $B_0=\bar\Sigma_BV$ is a solution to Eq.\ref{eqn:bond_sing_suf0}. Therefore, $\mathcal{B}(U\bar\Sigma_A,\bar\Sigma_BV)$ is a minimal singularity form of $\mathcal{B}(A,B)$.

For a general tensor network, such as a PEPS, the existence of a minimal singularity form can be proved in a similar manner. However, we lack an efficient algorithm to find the exact minimal singularity form. Therefore, we may adopt an iteration update method to obtain a tensor network whose singularity is minimal against perturbation on any bond, by performing bond regulation and sweeping all bonds, as explained in Sec.\ref{sec:sing}.
We have shown that this procedure can dramatically reduce the singularity of the tensor network, and allow steadily optimize the energy via ITE.

\section{Calculation of Renyi Entropy\label{sec:ent}}

In this section we introduce the methods to calculate the R\'{e}nyi entropy of a tensor network state\cite{Cirac11}.

%\subsection{Definition of Pseudo-reduced Density Matrix}

Let $|\Psi\rangle$ be the ground state of a quantum system. If we divide the system into two parts denoted by $\mathcal L$ and $\mathcal R$, the state can be written by
\begin{equation}
	|\Psi\rangle=\sum_{s_{\mathcal L}s_{\mathcal R}}\Psi_{s_{\mathcal L}s_{\mathcal R}}|s_{\mathcal L}\rangle|s_{\mathcal R}\rangle \, ,
\end{equation}
where $s_{\mathcal L}$ and $s_{\mathcal R}$ are physical indices of $\mathcal{L}$ and $\mathcal{R}$ parts respectively.
The reduced density matrix with respect to part $\mathcal{L}$ is
\begin{equation}
	\rho_{\mathcal L} =\tr_{\mathcal R}(|\Psi\rangle\langle\Psi|)
=\sum_{s_{\mathcal L}s'_{\mathcal L}s_{\mathcal R}}\Psi_{s_{\mathcal L}s_{\mathcal R}}\Psi^*_{s'_{\mathcal L}s_{\mathcal R}}|s_{\mathcal L}\rangle\langle s'_{\mathcal L}| \, ,
\label{eq:reducedDM}
\end{equation}
i.e., $(\rho_{\mathcal L})_{s_{\mathcal L}s'_{\mathcal L}}=\sum_{s_{\mathcal R}}\Psi_{s_{\mathcal L}s_{\mathcal R}}\Psi^*_{s'_{\mathcal L}s_{\mathcal R}}$.
For simplicity, we write Eq.\ref{eq:reducedDM} as $\rho_{\mathcal L}=\Psi\times\Psi^\dagger$.

The ground state of the system can be written as,
\begin{equation}
	\Psi_{s_{\mathcal L}s_{\mathcal R}}=\sum_{b_{\rm aux}} (\Psi_{\mathcal L})_{s_{\mathcal L} b_{\rm aux}}(\Psi_{\mathcal R})_{b_{\rm aux}s_{\mathcal R}}
\end{equation}
where $\Psi_{\mathcal L}$ and $\Psi_{\mathcal R}$ are the sub-tensor networks of part ${\mathcal L}$ and ${\mathcal R}$, respectively, and $b_{\rm aux}$ are
the auxiliary bonds that link the two sub-tensor networks.
In the matrix form,
\begin{equation}
	\Psi=\Psi_{\mathcal L}\times \Psi_R \, .
\end{equation}

\begin{figure}[tp!]
	\centering
	\includegraphics[scale=0.54]{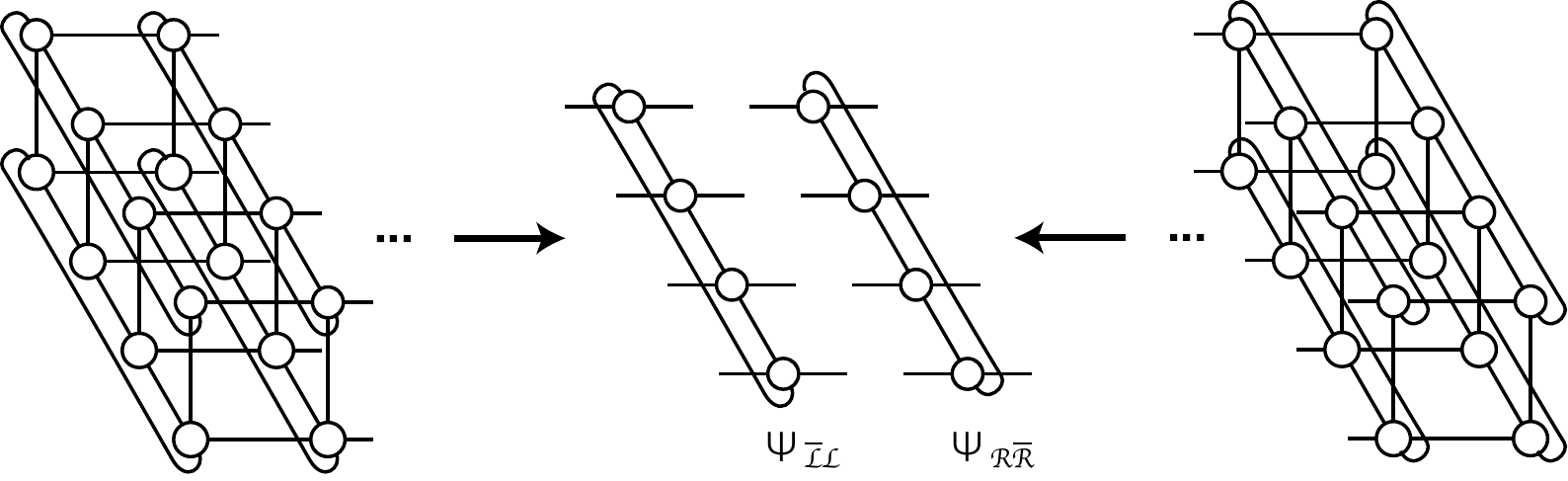}
	\caption{Schematic illustrations of $\Psi_{\bar {\mathcal L}{\mathcal L}}$ and $\Psi_{{\mathcal R}\bar {\mathcal R}}$.}
	\label{fig:lr_env}
\end{figure}

\begin{figure}[htp!]
	\centering
	\includegraphics[scale=0.54]{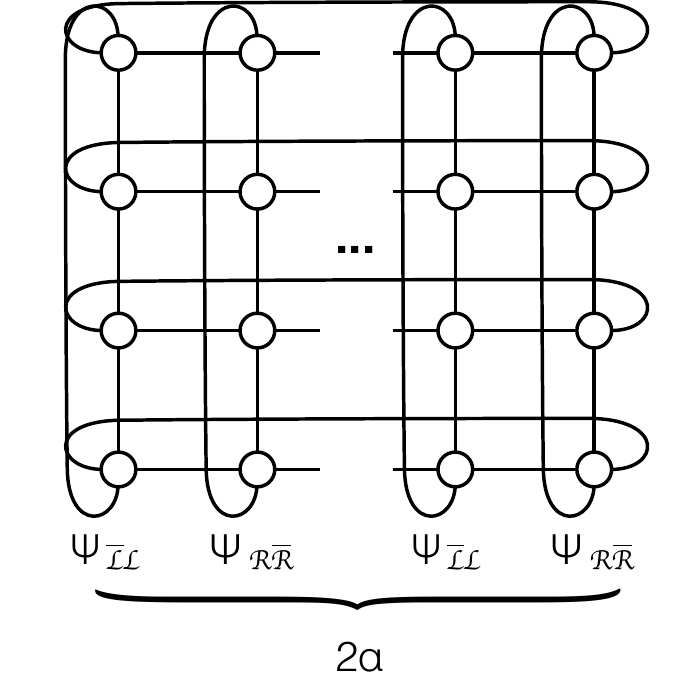}
	\caption{Contracting the tensor network $(\Psi_{\bar {\mathcal L}{\mathcal L}}\times\Psi_{{\mathcal R}\bar {\mathcal R}})^\alpha$ gives $\tr(\rho^\alpha_{\mathcal L})$.}
	\label{fig:density_trace}
\end{figure}

Then the trace of the $\alpha$-th power of the reduced density matrix can be obtained by,
\begin{align}
	\tr(\rho_{\mathcal L}^\alpha)&=\tr((\Psi\times\Psi^\dagger)^\alpha) \nonumber \\
	&=\tr((\Psi_{\mathcal L}\times \Psi_{\mathcal R}\times \Psi_{\mathcal R}^\dagger\times \Psi_{\mathcal L}^\dagger)^\alpha) \nonumber \\
	&=\tr((\Psi_{\mathcal L}^\dagger\times \Psi_{\mathcal L}\times \Psi_{\mathcal R}\times \Psi_{\mathcal R}^\dagger)^\alpha) \, .
\label{eq:N-DM}
\end{align}

For simplicity of the notation, we define $\Psi_{\bar {\mathcal L}{\mathcal L}}=\Psi_{\mathcal L}^\dagger\times \Psi_{\mathcal L}$ and $ \Psi_{{\mathcal R}\bar {\mathcal R}}= \Psi_{\mathcal R}\times \Psi_{\mathcal R}^\dagger$,
where  $(\Psi_{\bar {\mathcal L}{\mathcal L}})_{b'_{\rm aux}b_{\rm aux}}$=$\sum_{s_{\mathcal L}} (\Psi_{\mathcal L})^*_{s_{\mathcal L} b'_{\rm aux}}(\Psi_{\mathcal L})_{s_{\mathcal L} b_{\rm aux}}$ and $(\Psi_{{\mathcal R}\bar {\mathcal R}})_{b_{\rm aux}b'_{\rm aux}}$=$\sum_{s_{\mathcal R}} (\Psi_{\mathcal R})_{b_{\rm aux}s_{\mathcal R}}(\Psi_{\mathcal R})^*_{b'_{\rm aux}s_{\mathcal R}}$. We may calculate $\Psi_{\bar {\mathcal L}{\mathcal L}}$ and $\Psi_{{\mathcal R}\bar {\mathcal R}}$ \cite{lubasch14_2} as illustrated in Fig.\ref{fig:lr_env},
and $\tr(\rho_{\mathcal L}^\alpha)=\tr[(\Psi_{\bar {\mathcal L}{\mathcal L}}\times\Psi_{{\mathcal R}\bar {\mathcal R}})^\alpha]$, can be calculated by contracting the tensor network shown in Fig. \ref{fig:density_trace}.
The $\alpha$-th R\'{e}nyi entropy of between ${\mathcal L}$ and ${\mathcal R}$ parts of the system can be
calculated by Eq.~\ref{eqn:renyi_ent}.
Here, we only calculate the second order  R\'{e}nyi entropy (i.e., $\alpha$=2).

%\subsection{Calculation of 2nd Renyi Entropy Distribution in a Subspace \label{sec:any_entroty}}
\begin{figure}[t!]
	\centering
	\includegraphics[scale=0.6]{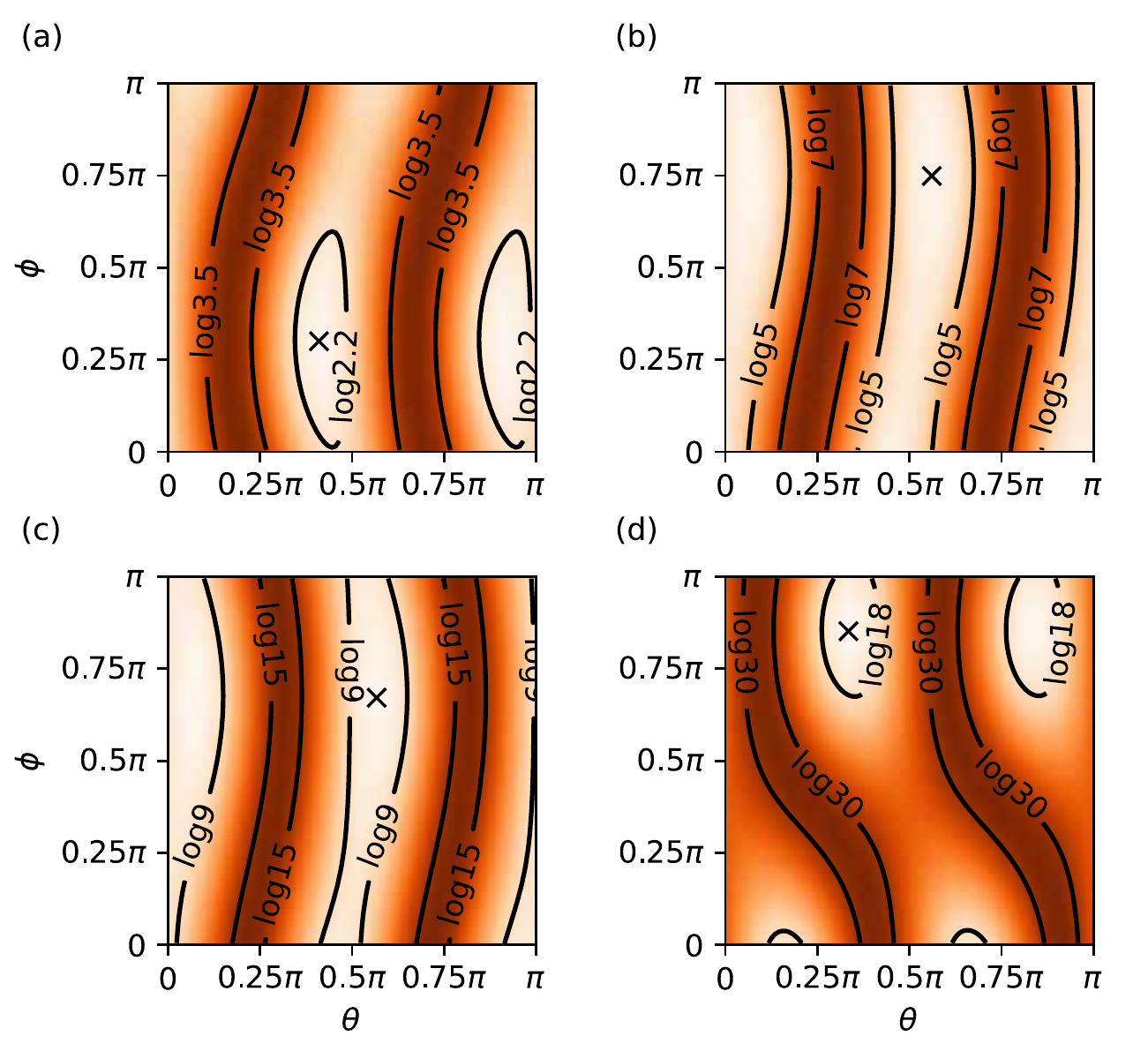}
	\caption{The second order R\'{e}nyi entropy $H_2$ of the toric code model with CBC, in GSM. The MES points are marked by ``$\times$''. The results are obtained on the (a) $4\times 20$, (b) $6\times 20$, (c) $8\times 20$ and (d) $10\times 20$ lattices.}
	\label{fig:toric_ent_scan}
\end{figure}

If the GSM of the system are $n$-fold degenerated, we first obtain $n$ linearly independent ground states $|\Psi_1\rangle,\dots |\Psi_n\rangle$ by independent simulations.
We then perform the Gram-Schmidt orthogonalization  process
to obtain the orthonormalized states of the GSM, $|e_1\rangle,\dots,|e_n\rangle$, where $|e_i\rangle=\sum_j v_{ij} |\Psi_j\rangle$.
The R\'{e}nyi entropy of MES is obtained by finding the minimal value of R\'{e}nyi entropy of $|\Psi({\bf c})\rangle=\sum_i c_i|e_i\rangle$ in the parameter space of ${\bf c}$.

Let $|\Psi({\bf c})\rangle=\sum_i c_i|e_i\rangle=\sum_{i=1}^n \bar c_i|\Psi_i\rangle$, where $\bar c_i=\sum_i c_iv_{ij}$,
be an arbitrary ground state, it is easy to show that,
\begin{align}
	\tr(\rho_{\Psi({\bf c})}^2)&=\tr(\Psi({\bf c})\times\Psi^\dagger({\bf c})\times\Psi({\bf c})\times\Psi^\dagger({\bf c})) \nonumber \\
	&=\sum_{ijkl} \bar c_i \bar c^\dagger_j \bar c_k \bar c^\dagger_l H_{ijkl}\, ,
\label{eqn:renyi_com}
\end{align}
where $H_{ijkl}=\tr(\Psi_i\times\Psi^\dagger_j\times\Psi_k\times\Psi^\dagger_l)$,
which can be calculated using the method discussed in the previous paragraphs.

Making use of the symmetry $H_{ijkl}=H_{klij}$ and $H_{ijkl}=H_{lkji}^*$, the number of independent $H_{ijkl}$ can be reduced from $n^4$ to $\frac 14 n^4+\frac 34n^2$. With these $H_{ijkl}$ calculated, one can obtain $\tr(\rho_{\Psi({\bf c})}^2)$, and thus the 2nd order R\'{e}nyi entropy using Eq.\ref{eqn:renyi_com}.

For the toric code model with CBC, the ground states are two-fold degenerate. Any ground state can be parameterized by $|\Psi(\theta,\phi)\rangle=\cos\theta |e_1\rangle +e^{i\phi}\sin\theta |e_2\rangle$. Using the above method, the R\'{e}nyi entropies $H_2$ in the $(\theta,\phi)$ plane for different cylinder widthes are calculated and shown in  Fig.\ref{fig:toric_ent_scan}(a)-(d), where the state with minimal $H_2$ in each figure corresponds to the MES.

%\bibliographystyle{apsrev4-2}
%\bibliography{mybib}% Produces the bibliography via BibTeX.

%apsrev4-2.bst 2019-01-14 (MD) hand-edited version of apsrev4-1.bst
%Control: key (0)
%Control: author (72) initials jnrlst
%Control: editor formatted (1) identically to author
%Control: production of article title (-1) disabled
%Control: page (0) single
%Control: year (1) truncated
%Control: production of eprint (0) enabled
%

\end{document}